\documentclass[aps,prb,notitlepage,twocolumn,10pt]{revtex4-1}
\usepackage{graphicx}
\usepackage{bm}
\usepackage[colorlinks=true]{hyperref}
\usepackage{xcolor}
\usepackage{amsmath,amssymb} 
\newcommand*{\nolink}[1]{\begin{NoHyper}#1\end{NoHyper}}

\begin{document}

\title{Ferromagnetic domain wall as a nonreciprocal string}

\author{Shu Zhang}
\author{Oleg Tchernyshyov}
\affiliation{
Department of Physics and Astronomy, 
Johns Hopkins University,
Baltimore, Maryland 21218, USA
}
\date{\today}

\begin{abstract}
We present a simple model of a domain wall in a thin-film ferromagnet. A domain wall is represented as a nonreciprocal string, on which transverse waves propagate with different speeds in opposite directions. The model has three parameters: mass density, tension, and a gyroscopic constant quantifying the nonreciprocity. We discuss the unusual dynamics of a nonreciprocal string in finite geometry. It agrees well with numerically simulated motion of a ferromagnetic domain wall in a strip of constant width. 
\end{abstract}

\maketitle

\section{Introduction}

\begin{figure}[t]
\includegraphics[width=0.95\columnwidth]{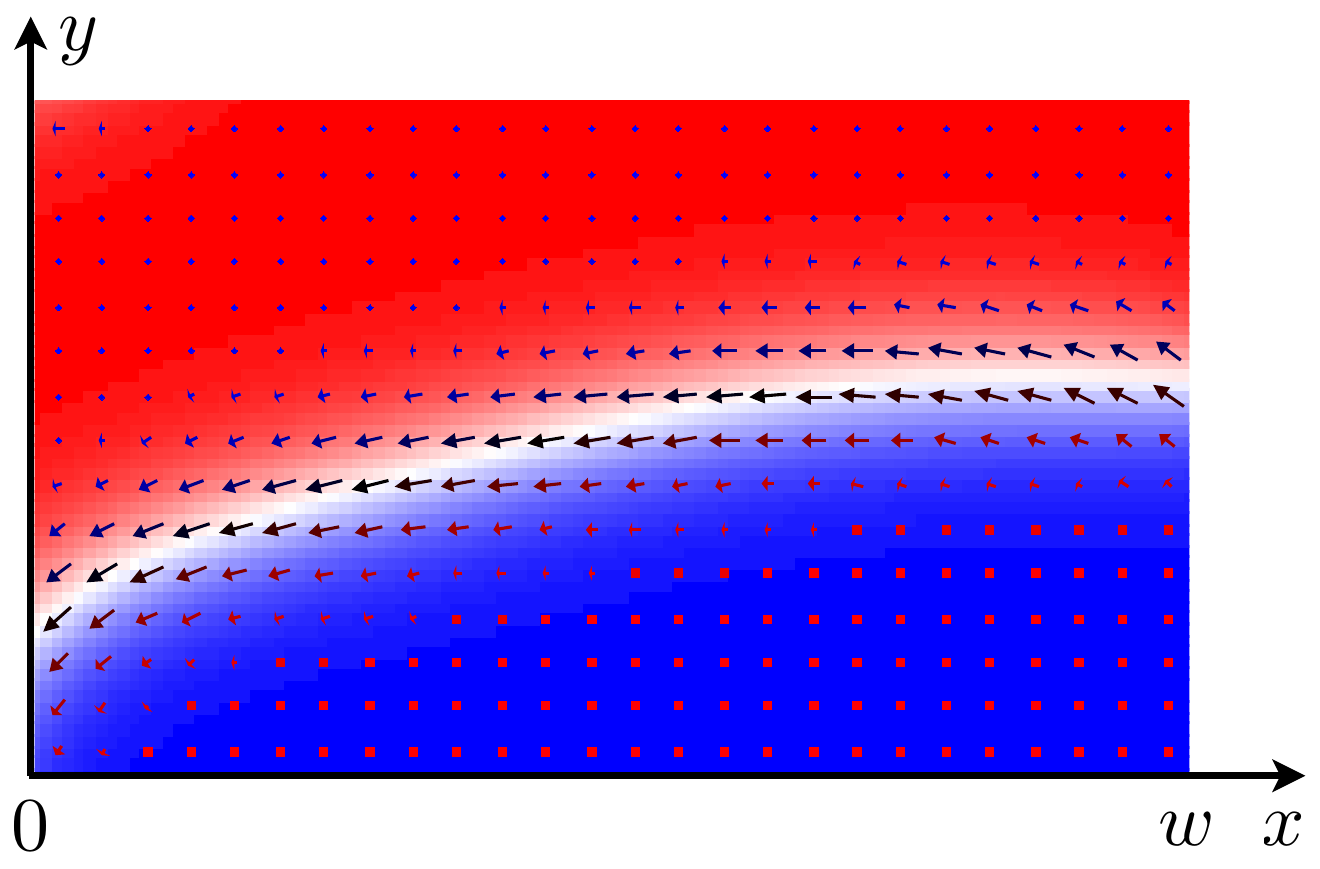}
\caption{Domain wall in a ferromagnetic strip $0 \leq x \leq w$ with an easy axis $\hat{\mathbf z}$. Colors encode $m_z$: red is $m_z = +1$, blue is $m_z = -1$, and white is $m_z = 0$. Arrows indicate in-plane components $m_x$ and $m_y$.}
\label{fig:domain-wall}
\end{figure}

The goal of this paper is to present a simple dynamical model of a domain wall in a thin ferromagnetic film. A domain wall is the boundary separating two domains of uniform magnetization (Fig.~\ref{fig:domain-wall}). In two dimensions, a boundary is a line. Therefore, even the simplest dynamical model must treat a domain wall as an extended object with infinitely many degrees of freedom.  

Our approach has been inspired by the success of simple dynamical models of point-like magnetic solitons such as the domain wall in one spatial dimension \cite{Walker1974} and the vortex in two.\cite{Huber1982} Although both a domain wall and a vortex have internal structure, the response to weak external perturbations is dominated by the soft modes of global translations and rotations. 

For example, Thiele's equation 
\begin{equation}
\mathbf G \times \dot{\mathbf R} - \partial U/ \partial \mathbf R - D \dot{\mathbf R} = 0 
\end{equation}
for a vortex center $\mathbf R$ expresses the dynamical equilibrium between the intrinsic gyroscopic force, conservative force (e.g., from an applied magnetic field), and viscous force, respectively. \cite{Thiele1973} The velocity-dependent gyroscopic force $\mathbf G \times \dot{\mathbf R}$ is similar in nature to the Lorentz force acting on an electric charge in a magnetic field and to the Coriolis force acting on a massive object in a rotating frame. Such forces break the symmetry of time reversal; their magnitudes are proportional to spontaneous magnetization, magnetic field, and rotation frequency of the reference frame, respectively. 

In a similar way, we seek the simplest coarse-grained description of a domain wall modeled as a line in two spatial dimensions. We deem essential three ingredients:

\emph{Tension.} A domain wall is an excitation raising the energy of the system. In a model with local interactions, it is natural to expect an energy cost proportional to the length $\ell$ of the domain wall, $E = \sigma \ell$. This defines line tension $\sigma$.

\emph{Inertia.} A domain wall in a ferromagnet may possess inertia quantified by the D{\"o}ring mass \cite{Doring1948,Malozemoff1979} proportional to the domain-wall length, $M_D = \rho \ell$, where $\rho$ is the linear mass density.

\emph{Gyroscopic effect.} The breaking of the time reversal symmetry by spontaneous magnetization strongly impacts the dynamics of a domain wall. The effect is quantified by a gyroscopic constant $g$ proportional to the spontaneous magnetization $\mathcal M$ of the ferromagnet. 

The first two ingredients yield the familiar wave equation $ \rho \ddot{y} - \sigma y'' = 0$, whose solutions are waves propagating in both directions with the speed $v_0 = \sqrt{\sigma/\rho}$. The third ingredient introduces a simple modification of the wave equation: 
\begin{equation}
\rho \ddot{y} - \sigma y'' + 2 g \dot{y}' = 0,
\label{eq:eom-y-intro}
\end{equation}
The mixed derivative $\dot{y}'$ violates the symmetries of time reversal $t \mapsto -t$ and mirror reflection $x \mapsto - x$. It is therefore forbidden for a regular string, which respects both symmetries. In a ferromagnet, both symmetries are spontaneously broken and the mixed term is allowed.  

The modified wave equation (\ref{eq:eom-y-intro}) has waves propagating left and right with different speeds,
\begin{equation}
v_\pm = \pm v_0 e^{\pm u}.
\end{equation}
The nonreciprocity of wave propagation is quantified by the dimensionless parameter 
\begin{equation}
\sinh{u} = \frac{g}{\sqrt{\rho \sigma}}.
\end{equation}
In thin-film ferromagnets, $u$ can be of order 1, so that the dynamics of a domain wall is strongly nonreciprocal. This motivates us to take a close look at the model of a nonreciprocal string. We shall see that it has rather unusual normal modes and that its mechanics is strikingly different from that of a regular string. 

Because spatial and temporal derivatives enter the wave equation (\ref{eq:eom-y-intro}) in a homogeneous way, it makes sense to treat time $t$ and spatial coordinate $x$ on an equal footing, as coordinates in spacetime $(t,x) \equiv (x^0,x^1)$. Then Eq.~(\ref{eq:eom-y-intro}) reads $\Lambda^{\mu\nu} \partial_\mu \partial_\nu y = 0$, where $(\partial_0,\partial_1) \equiv (\partial_t,\partial_x)$ and 
\begin{equation}
\Lambda^{\mu\nu} = 
	\left(
    	\begin{array}{cc}
    		\rho & g \\
            g & -\sigma
    	\end{array}
    \right).
\end{equation}

The Lagrangian of a freely moving string is 
\begin{equation}
\mathcal L 
	= \frac{1}{2} \Lambda^{\mu\nu} 
    	\partial_\mu y \, \partial_\nu y.
\end{equation}
Weak perturbations can be added as terms linear in the displacement field $y$ and its derivatives $\partial_\mu y$. The perturbations are quite generally represented by a gauge field $a_\mu$ and by a source field $p$: 
\begin{equation}
\mathcal L 
	= \frac{1}{2} \Lambda^{\mu\nu} 
    	(\partial_\mu y - a_\mu) 
        (\partial_\nu y - a_\nu) 
        + py.
\end{equation}
The source field $p$ expresses pressure (force per unit length) exerted on the string and comes from perturbations such as a magnetic field normal to the film plane. The temporal component of the gauge field $a_0 = - \tau/g$ is proportional to the torque density (per unit length) $\tau$ from a magnetic field or spin-polarized current in the film plane. We have not found physical perturbations generating the spatial component of the gauge field $a_1$.

Our minimal model excludes the effects of a stray magnetic field and the bending energy of a domain wall. These omissions can be justified in certain limits (thin film, long-wavelength deformations) and greatly simplify the mathematical analysis. Furthermore, the model may be applicable to other string-like objects with a broken time-reversal symmetry. Therefore the mechanics of a nonreciprocal string presents interests beyond magnetism. 

The paper is organized as follows. In Sec.~\ref{sec:model}, we present the phenomenological model of a nonreciprocal string and characterize its physical properties. In Sec.~\ref{sec:derivation}, we derive this phenomenology from a micromagnetic model of a N{\'e}el domain wall and discuss the range of its applicability. We apply the phenomenological model to deduce the response of a ferromagnetic domain wall to weak external perturbations in Sec.~\ref{sec:response}. Numerical tests of the model via micromagnetic simulations are described in Sec.~\ref{sec:simulations}. Sec.~\ref{sec:discussion} contains concluding remarks and an outlook. Technical information is contained in appendices.

\section{Model of a nonreciprocal string}
\label{sec:model}

\subsection{Lagrangian and equations of motion}

We begin with a phenomenological justification for the model of a nonreciprocal string described as a line in a two-dimensional space $(x,y)$. We will assume that the string can be parametrized as a function $y(x)$ with $0 \leq x \leq w$ and $-\infty < y < +\infty$. To describe the dynamics of the string, we must add the time variable, hence $y(t,x)$. We thus treat $y$ as a field living in $(1+1)$-dimensional spacetime
\begin{equation}
\Omega = (t,x): 
\quad
-\infty < t < +\infty,
\quad 
0 \leq x \leq w. 
\label{eq:Omega}
\end{equation}
Here $w$ is the length of the string in equilibrium, when $y(t,x) = Y = \mathrm{const}$. 

Potential energy of the string is proportional to its length, 
\begin{equation}
U = \sigma \int_0^w dx \, \sqrt{1+{y'}^2} 
	\approx \sigma w + \frac{\sigma}{2} \int_0^w dx \, {y'}^2
\end{equation}
for small deviations from equilibrium, $y' \ll 1$. Here $\sigma$ is the string tension. Kinetic energy of transverse vibrations is $\int_0^w dx \, \rho \dot{y}^2/2$, where $\rho$ is the linear mass density. These ingredients define the Lagrangian of a regular string,
\begin{equation}
\mathcal L =  
		\frac{\rho \dot{y}^2}{2} 
    	- \frac{\sigma {y'}^2}{2}.
\label{eq:L-reciprocal}
\end{equation}
Minimization of the action $S = \int_\Omega dt \, dx \, \mathcal L$ yields the familiar wave equation $ \rho \ddot{y} - \sigma y'' = 0$ and the boundary conditions $y'(t,0) = y'(t,w) = 0$ for a string with free ends.

We wish to extend the regular string model by adding some terms that preserve the linearity of the equations of motion and translational symmetry in the transverse ($y$) direction. We are thus allowed to add terms at most quadratic in the derivatives $\dot{y}$ and $y'$ to the Lagrangian (\ref{eq:L-reciprocal}). With $\dot{y}^2$ and ${y'}^2$ already present, the only new quadratic term would be $\dot{y} y'$. (Linear terms, representing external perturbations, will be dealt with later.) Hence 
\begin{equation}
\mathcal L =  
		\frac{\rho \dot{y}^2}{2} 
    	- \frac{\sigma {y'}^2}{2} 
    	+ g \dot{y} y'.
\label{eq:L-nonreciprocal}
\end{equation}
The coupling constant $g$ of the new term has the dimension of angular momentum per unit area. As we shall see later, this term is ultimately related to the precessional character of magnetization dynamics in a ferromagnet. For this reason, we call this term \emph{gyroscopic}.

The gyroscopic term $g \dot{y} y'$ in Eq.~(\ref{eq:L-nonreciprocal}) breaks the symmetries of time reversal $t \mapsto -t$ and mirror reflection $x \mapsto -x$ and is therefore forbidden for a regular string. In a ferromagnet, both of these symmetries are broken and this term is allowed. 

Minimization of the action $S = \int_\Omega dt \, dx \, \mathcal L$ yields the equation of motion in the bulk (\ref{eq:eom-y-intro}),
\begin{equation}
\rho \ddot{y} - \sigma y'' + 2 g \dot{y}' = 0,
\label{eq:eom-y}
\end{equation}
and the boundary conditions for a string with free ends,
\begin{equation}
\sigma y' - g \dot{y} = 0 
\mbox{ for }
x = 0, w.
\label{eq:bc-y}
\end{equation}
The presence of a mixed derivative $\dot{y}'$ in the wave equation (\ref{eq:eom-y}) makes the propagation of waves nonreciprocal. Therefore, the velocities $v_{\pm}$ of right and left-moving waves differ not only in sign but also in magnitude: 
\begin{equation}
\omega = v_{\pm} k,
\quad
v_{\pm} 
	= \frac{g}{\rho} \pm \sqrt{\frac{g^2}{\rho^2} + \frac{\sigma}{\rho}}
	= \pm v_0 e^{\pm u}. 
\label{eq:v}
\end{equation}
Here $v_0 = \sqrt{\sigma/\rho}$ is the propagation speed for a regular string ($g = 0$) and $\sinh{u} = g/\sqrt{\rho \sigma}$ is a dimensionless measure of nonreciprocity.

\subsection{Normal modes of a finite string}

To find the normal modes of a finite string, $y_n(x,t) = f_n(x) e^{- i \omega_n t}$, we first note that an isolated pulse propagating along the domain wall and bouncing off the string ends at $x=0$ and $w$ makes a round trip in the time 
\begin{equation}
T = \frac{w}{v_+} - \frac{w}{v_-} = \frac{2w}{c},
\label{eq:T}
\end{equation}
where the speed $c$ is defined as
\begin{equation}
\frac{1}{c} 
	= \sqrt{\frac{g^2}{\sigma^2} + \frac{\rho}{\sigma}} 
	=  \frac{\cosh{u}}{v_0}.
\label{eq:c}
\end{equation}
Hence the eigenfrequencies 
\begin{equation}
\omega_n = \frac{2\pi n}{T} = \frac{n \pi c}{w},
\quad
n = 1, 2, 3, \ldots
\label{eq:omega-n}
\end{equation}
The normal modes are superpositions of right and left-moving waves with the wavenumbers
\begin{equation}
k_{n\pm} = \frac{\omega_n}{v_\pm} = \pm \frac{2\pi n}{w(e^{\pm 2u}+1)}.
\label{eq:dispersion}
\end{equation}
For free boundary conditions (\ref{eq:bc-y}), the normal modes are
\begin{eqnarray}
y_n(t,x) &=& a_n e^{- i \omega_n t}(e^{i k_{n+}x} + e^{i k_{n-}x})/2 
\nonumber\\
	&=& a_n \cos{(\pi n x/w)} 
    	e^{- i \pi n (ct + x \tanh{u})/w}.
\label{eq:normal-modes}
\end{eqnarray}
Normal modes of a string with fixed ends are
\begin{eqnarray}
y_n(t,x) &=& a_n e^{- i \omega_n t}(e^{i k_{n+}x} - e^{i k_{n-}x})/2i 
\nonumber\\
	&=& a_n \sin{(\pi n x/w)} 
    	e^{- i \pi n (ct + x \tanh{u})/w}.
\label{eq:normal-modes-fixed-ends}
\end{eqnarray}

Because of a mismatch $k_{n+} \neq - k_{n-}$, the normal modes are not simply standing waves but include a running component under the familiar standing-wave envelope (see animations in Supplemental Material \cite{suppmat}). For $g \neq 0$, modes with different $n$ are not mutually orthogonal. The nonorthogonality can be traced to the eigenproblem $- \sigma f'' - 2 i \omega g f' = \rho \omega^2 f$ not being of the Sturm-Liouville type for $g \neq 0$: the operator $- \sigma \, \partial_x^2 - 2 i \omega g \, \partial_x$ depends on the eigenfrequency $\omega$.  

\subsection{Zero mode}

In addition to periodic normal modes, a string with free ends has a zero mode associated with the translational symmetry in the $y$ direction,
\begin{equation}
y_0(t,x) = a_0 (ct + x \tanh{u}).
\label{eq:zero-mode}
\end{equation}
Linear proportionality between the velocity and tilt,
\begin{equation}
y_0' = \frac{g}{\sigma} \dot{y}_0
	= \frac{1}{c}\dot{y}_0 \tanh{u},
\label{eq:tilt-velocity}
\end{equation}
is dictated by the boundary conditions (\ref{eq:bc-y}). 

Because normal modes are of oscillatory nature and share a common period $T$, the time-averaged transverse velocity of the string is determined by the zero mode alone.

\subsection{Motion of a tilted-and-released string}

\begin{figure}[t]
\includegraphics[width=\columnwidth]{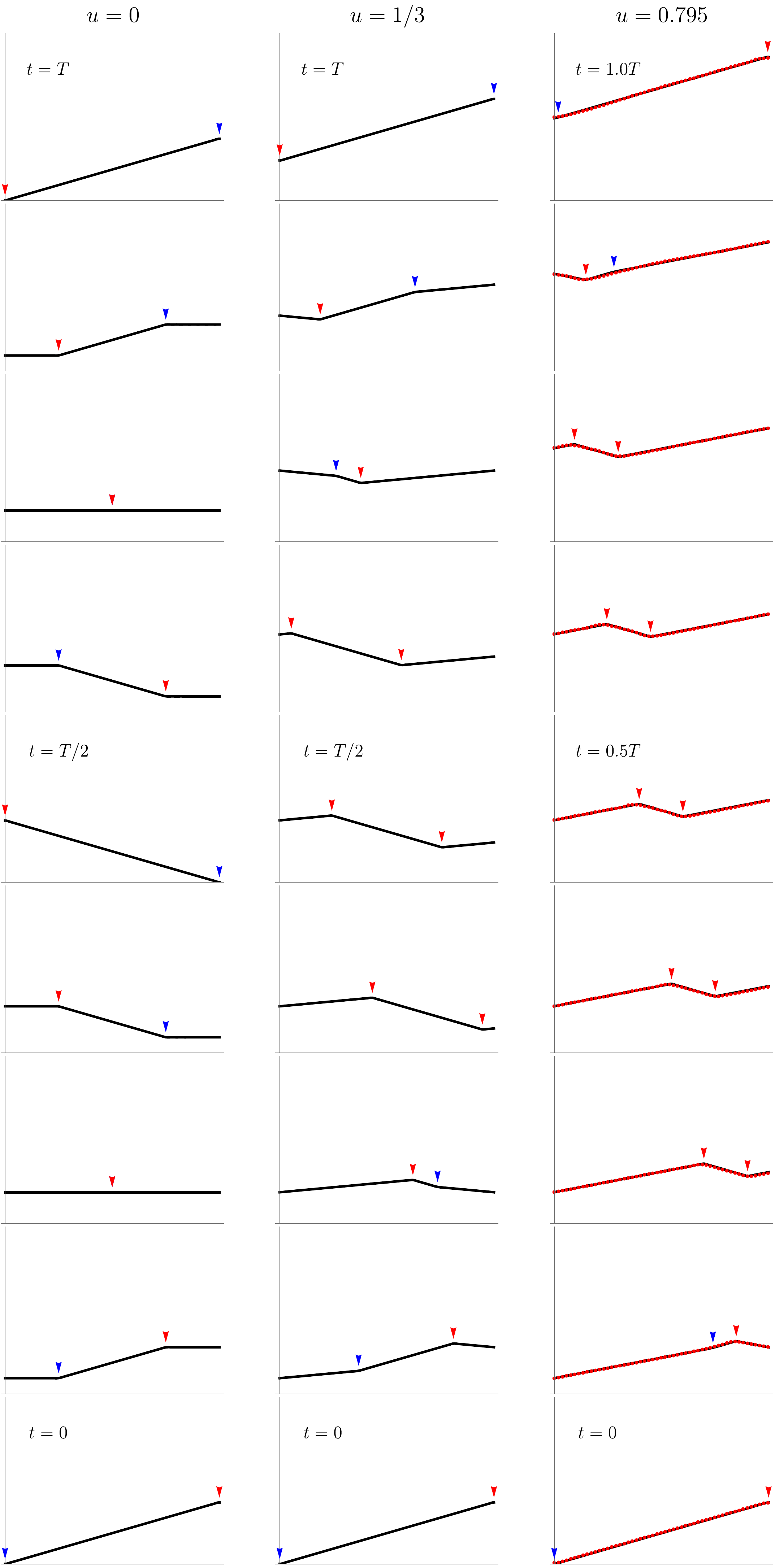}
\caption{Time evolution from $t=0$ (bottom row) to $T$ (top row) for a tilted-and-released string with nonreciprocity $u = 0$ (left column), 1/3 (central column), and 0.795 (right column). Lines represent theory, dots (right column) are from micromagnetic simulations. Blue and red arrows mark right and left-moving kinks, respectively. See Supplemental Material \cite{suppmat} for animations.}
\label{fig:snapshots}
\end{figure}

A nonreciprocal string responds in strikingly unusual ways to external perturbations. Consider a string that is uniformly tilted, $y'(0,x) = a$, and released with zero initial velocity, $\dot{y}(0,x) = 0$ (see Fig.~\ref{fig:snapshots} and its animation in Supplemental Material \cite{suppmat}). A regular string ($u = 0$) oscillates periodically, with two kinks propagating back and forth on it. A nonreciprocal string ($u \neq 0$) also moves with an average velocity proportional to the initial tilt, 
\begin{equation}
\frac{1}{T} \int_0^T dt \, \dot{y}(t,x)
	\equiv \langle\dot{y}(x)\rangle_t 
    = c a \tanh{u}.
\label{eq:average-velocity-kinks}
\end{equation}

This can be seen as follows. At $t = 0$, two kinks emerge from the edges and move into the bulk, dividing the string into three straight segments. The motion of the outward segments is described by the zero mode (\ref{eq:zero-mode}). Their transverse velocities $\dot{y} = \pm ca$ and slopes $y' = \pm a \tanh{u}$ are determined by the tilt-velocity relation (\ref{eq:tilt-velocity}) and the longitudinal kink velocities $v_{\pm}$ (\ref{eq:v}). As the right-moving kink reaches the right edge at $t = w/v_+$, the right end of the string switches its transverse velocity from $-ca$ to $+ca$ for the remainder of the period $T$ (\ref{eq:T}). Averaging the velocity over the period yields Eq.~(\ref{eq:average-velocity-kinks}) for $x = w$. And as the shape of the string is periodic in time, the time-averaged velocity is the same everywhere, $\langle\dot{y}(x)\rangle_t = \langle\dot{y}\rangle_t$.

Our attempts to obtain the time evolution of a nonreciprocal string for general initial conditions have been frustrated by a lack of orthogonality relations for normal modes (\ref{eq:normal-modes}). Although one can expand an arbitrary motion in terms of the normal modes, finding their amplitudes is a nontrivial problem. Below we focus on deriving the time-averaged transverse velocity $\langle\dot{y}\rangle_t$, a quantity most easily accessible in experiments and associated with the zero mode (\ref{eq:zero-mode}). A general solution of the initial-value problem via the Laplace transform is provided in Appendix \ref{app:Laplace}.

\subsection{Transverse momentum}

To that end, we start with the transverse momentum $P_y$, a conserved quantity by virtue of translational invariance in the $y$ direction. It is convenient to viewed global translations as a field transformation 
\begin{equation}
\tilde{y}(t,x) = y(t,x) + Y 
\label{eq:translation-global}
\end{equation}
in two-dimensional spacetime $(x^0, x^1) \equiv (t,x)$. Application of Noether's theorem yields a conserved current, $\partial_\mu j^\mu = 0$, where $(\partial_0, \partial_1) \equiv (\partial_t, \partial_x)$, with components
\begin{equation}
j^0 = \frac{\partial \mathcal L}{\partial \dot{y}}
	= \rho \dot{y} + g y',
\quad
j^1 = \frac{\partial \mathcal L}{\partial y'}
	= - \sigma y' + g \dot{y}.
\label{eq:j}
\end{equation}
The global conserved charge is obtained by integrating the charge density $j^0$ over space, 
\begin{equation}
P_y = \int_0^w dx \, (\rho \dot{y} + g y').
\label{eq:Py-def}
\end{equation}
Note that transverse momentum (\ref{eq:Py-def}) has, in addition to the obvious kinetic part $P_y^\mathrm{kin} = \int_0^w dx \, \rho \dot{y}$, a geometric contribution 
\begin{equation}
P_y^\mathrm{geom} = \int_0^w dx \, g y' 
	= g [y(t,w)-y(t,0)].
\end{equation} 
It is a common feature of ferromagnetic solitons.\cite{Thiele1976, Haldane1986, Papanicolaou1991, Tchernyshyov2015}

By linearity, momentum $P_y$ is a sum of momenta carried by the normal modes. It can be checked directly that the contribution of any periodic mode (\ref{eq:normal-modes}) is zero. Therefore, all of $P_y$ comes from the zero mode (\ref{eq:zero-mode}), whose amplitude determines the time-averaged velocity $\langle \dot{y} \rangle_t$. We thus expect a linear relation 
\begin{equation}
P_y = M_\mathrm{eff} \langle \dot{y} \rangle_t,
\label{eq:Py-dot-y}
\end{equation}
where the effective mass $M_\mathrm{eff}$ remains to be determined. This is most easily done in a state where only the zero mode (\ref{eq:zero-mode}) is excited. The string moves at a constant velocity $\dot{y} = a_0 c$ and carries momentum $P_y = a_0 w(\rho c + g \tanh{u})$. Hence the effective mass  
\begin{equation}
M_\mathrm{eff} = (\rho + g^2/\sigma) w 
	\equiv \rho_\mathrm{eff} w
\label{eq:M}
\end{equation}
that is different from the D{\"o}ring mass $M_D = \rho w$. 
The effective mass (\ref{eq:M}) defines the response to an external force $F$, which can be generated, e.g., by the application of an external field along the easy axis. The string accelerates with the time-averaged acceleration $\langle \ddot{y} \rangle_t = F/M_\mathrm{eff}$.

Thus, given an arbitrary initial configuration of the string $y(0,x)$ and initial velocity $\dot{y}(0,x)$, we can compute the conserved momentum (\ref{eq:Py-def}) and use Eqs.~(\ref{eq:Py-dot-y}) and (\ref{eq:M}) to determine the average velocity $\langle \dot{y} \rangle_t$ of the string. 

\subsection{Covariant formalism}

It is instructive to analyze the string dynamics in a covariant fashion as in theory of relativity. Write the Lagrangian in the tensor notation,
\begin{equation}
\mathcal L 
	= \frac{1}{2}\Lambda^{\mu\nu} 
    	\partial_\mu y \, \partial_\nu y,
\label{eq:L-tensor-notation}
\end{equation}
where 
\begin{equation}
\Lambda^{\mu\nu} 
	= \left(
        	\begin{array}{cc}
            	\rho & g \\
                g & - \sigma
            \end{array}
        \right)
	= \frac{\sigma}{c^2} 
		\left(
    		\begin{array}{cc}
    		 	\mathop{\mathrm{sech}^2{u}} 
            		& c \tanh{u} \\
            	c \tanh{u} & -c^2
    		\end{array}
    	\right).
\label{eq:metric}
\end{equation}
Equation of motion (\ref{eq:eom-y}) reads 
\begin{equation}
\Lambda^{\mu\nu} \partial_\mu \partial_\nu y = 0.
\end{equation}
The boundary condition for a string with free ends (\ref{eq:bc-y}) becomes
\begin{equation}
\Lambda^{1\nu} \partial_\nu y = 0
\mbox{ for } x = 0, w.
\end{equation}
The conserved current (\ref{eq:j}) is 
\begin{equation}
j^\mu
	= \frac{\partial \mathcal L}
		{\partial (\partial_\mu y)}
    = \Lambda^{\mu\nu} \partial_\nu y.
\label{eq:j-covariant}
\end{equation}

A transformation to a local time $(\tilde{t}, \tilde{x})$, where 
\begin{equation}
\tilde{t} = t + (x/c) \tanh{u}, 
\quad 
\tilde{x} = x, 
\end{equation}
brings the Lagrangian to a diagonal form with 
\begin{equation}
\tilde{\Lambda}^{\mu\nu} 
	= \left(
        	\begin{array}{cc}
            	\tilde{\rho} & \tilde{g} \\
                \tilde{g} & - \tilde{\sigma}
            \end{array}
        \right)
    = \frac{\sigma}{c^2} 
		\left(
    		\begin{array}{cc}
    			1 & 0 \\ 
            	0 & -c^2
    		\end{array}
    	\right).
\end{equation}
In this frame, the string has tension $\tilde{\sigma} = \sigma$, mass density $\tilde{\rho} = \rho + g^2/\sigma$, and is reciprocal, $\tilde{g} = 0$: right and left-moving waves have equal speeds, $\tilde{v}_\pm = \pm c$. From the transformation of the derivatives,
\begin{equation}
\partial_t = \partial_{\tilde{t}}, 
\quad
\partial_x = \partial_{\tilde{x}} + (1/c) \partial_{\tilde{t}} \tanh{u},
\end{equation}
we find that the free boundary conditions (\ref{eq:bc-y}) read $\partial_{\tilde x}y = 0$ at $\tilde{x} = 0$, $w$ and yield the familiar normal modes of a regular string, $y_{n}(\tilde{t}, \tilde{x}) = \cos{(\pi n \tilde{x}/w)} e^{- i \pi n c \tilde{t}/w}$. Returning to the global-time frame $(t,x)$ gives Eq.~(\ref{eq:normal-modes}). 

\subsection{Background gauge field}
\label{subsec:gauge}

We will also find it useful to introduce a background gauge field $a_\mu$ coupled to the string displacement $y$: 
\begin{equation}
\mathcal L 
	= \frac{1}{2} \Lambda^{\mu\nu} 
    	(\partial_\mu y - a_\mu)
        (\partial_\nu y - a_\nu).
\label{eq:L-minimal-coupling}
\end{equation}
This coupling is similar to the one between the electromagnetic gauge field and the phase of a superconducting order parameter.\cite{Tinkham} The Lagrangian is manifestly covariant under gauge transformations 
\begin{equation}
\tilde{y} = y - \chi,
\quad
\tilde{a}_\mu = a_\mu - \partial_\mu \chi,
\label{eq:translation-local}
\end{equation}
where $\chi(t,x)$ is an arbitrary function. The gauge symmetry (\ref{eq:translation-local}) is merely a local version of global translations (\ref{eq:translation-global}). The conserved current associated with this gauge symmetry, 
\begin{equation}
j^\mu 
	= - \frac{\partial \mathcal L}{\partial a_\mu}
	= \Lambda^{\mu\nu} (\partial_\nu y - a_\nu),
\label{eq:j-gauge}
\end{equation}
is identical to the conserved current in Eq.~(\ref{eq:j-covariant}) if we set $a_\nu = 0$. 

The equation of motion for $y$,
\begin{equation}
\partial_\mu [\Lambda^{\mu\nu}(\partial_\nu y - a_\nu)]
	= 0,
\label{eq:eom-y-gauge}
\end{equation}
expresses conservation of current (\ref{eq:j-gauge}). The boundary condition for a free string, 
\begin{equation}
\Lambda^{1\nu}(\partial_\nu y - a_\nu) = 0
\mbox{ for } x=0, w,
\label{eq:bc-gauge}
\end{equation}
indicates the vanishing of the current at the string ends, $j^1(t,0) = j^1(t,w) = 0$.

The equation of motion of a string in a background gauge field (\ref{eq:eom-y-gauge}), still linear in $y$, acquires an inhomogeneous part due to the gauge field. Solving it in a general case is still a nontrivial problem because of the non-orthogonality of the normal modes (\ref{eq:normal-modes}). The situation simplifies if the gauge field is trivial, i.e., when its gauge-invariant curvature, the electric field $F_{01} = \partial_0 a_1 - \partial_1 a_0$, vanishes. Then $a_\mu$ is a pure gauge, 
\begin{equation}
a_\mu = \partial_\mu \chi.
\label{eq:a-pure-gauge}
\end{equation}
We can use the gauge transformation (\ref{eq:translation-local}) to relate a solution for a trivial gauge background $y$ to that of a free string, e.g., $\tilde{y} = 0$:
\begin{equation}
y = \tilde{y} + \chi = \chi.
\label{eq:response-to-pure-gauge}
\end{equation}
Thus, for a given trivial background $a_\mu$, a partial solution can be found by simply finding a gauge $\chi$ solving Eq.~(\ref{eq:a-pure-gauge}).

\subsection{Dissipation}

Viscous damping, neglected so far in our analysis, will gradually dissipate the momentum of a moving string and it will come to rest. To account for it, we add a viscous force $-b \dot{y}$ to the equation of motion (\ref{eq:eom-y}). For a domain wall, the viscosity coefficient $b = \alpha|g|/\lambda$,\cite{Clarke2008} where $\lambda$ is the width of the domain wall and $\alpha \ll 1$ is Gilbert's dimensionless damping constant.\cite{Gilbert2004} The inclusion of damping yields $\dot{P}_y = - bw \langle \dot{y} \rangle$, from which we obtain the relaxation rate for momentum and average velocity, 
\begin{equation}
\Gamma = (\alpha c/\lambda) |\tanh{u}|.
\end{equation}
The relaxation rate for periodic normal modes is $\Gamma/2$. Attenuation of the zero mode over one period is $e^{-\Gamma T} = e^{-(2\alpha w/\lambda)|\tanh{u}|}$. In insulating ferromagnets, Gilbert's damping  $\alpha$ can be as low as $10^{-4}$ in insulating ferromagnets.\cite{Klingler2017, Stasinopoulos2017} Therefore, even for a strip whose width $w$ greatly exceeds the domain-wall width $\lambda$, the natural modes (\ref{eq:normal-modes}) can be underdamped. Observing the peculiar dynamics of a nonreciprocal string may well be feasible.

\section{Derivation of the model}
\label{sec:derivation}

In this section, we obtain the model of a nonreciprocal string as a long-wavelength, low-frequency limit of a micromagnetic model of a domain wall in a thin-film ferromagnet. This was previously done for a Bloch domain wall.\cite{Makhfudz2012} Here we do it for a N{\'eel} domain wall stabilized by sufficiently strong Dzyaloshinskii-Moriya interaction.\cite{EPL.100.57002}

\subsection{Gyroscopic constant}

Whereas mass density $\rho$ and string tension $\sigma$ depend on the specifics of the micromagnetic model, the gyroscopic constant $g$ is model-independent and is fully determined by a single material parameter: the density of angular momentum per unit area $\mathcal J$. This becomes clear from the examination of the corresponding term in the action $S_g = g \int dt \, dx \, \dot{y} y'$. This part of the action is independent of how fast the string moves between its initial and final states and is thus a purely geometrical quantity that has nothing to do with the energetics of the underlying micromagnetic model. This geometrical action comes from the Berry phase of the spins making up the ferromagnet. 

In a ferromagnetic film of thickness $h$, $\mathcal J = h \mathcal M/\gamma$, where $\mathcal M$ is the length of magnetization (magnetic dipole moment per unit volume) and $\gamma$ is the gyromagnetic ratio. Spontaneous magnetization has values $\mathbf M = \pm \mathcal M \hat{\mathbf z}$ deep in the upper and lower domains of Fig.~\ref{fig:domain-wall}. The angular momentum per unit area is $\pm \mathcal J \hat{\mathbf z}$, respectively. Suppose the domain wall is moving at the velocity $\dot{y}_0$ with a uniform tilt $y_0'$ (only the zero mode is engaged). The motion of the domain boundary changes the angular momentum of the ferromagnet at the rate $d \mathbf J/dt = - 2 \mathcal J \hat{\mathbf z} w \dot{y}_0$. This change is brought by the torque applied to the domain wall by the edges. A tilted domain wall is lengthened and its energy is thus increased by $\sigma w {y_0'}^2/2$, creating torque $\bm \tau = - \sigma w y_0' \hat{\mathbf z}$. From $\bm \tau = d \mathbf J/dt$ we obtain a proportionality between velocity and tilt, $\sigma y_0' = 2 \mathcal J \dot{y}_0$. Comparison to the boundary conditions (\ref{eq:bc-y}) relates the gyroscopic constant to density of angular momentum, $g = 2 \mathcal J$.  

The gyroscopic coupling $g$ of a domain wall is a topological quantity in the sense that it depends only on the topology of the domain wall but not on its shape, detailed structure, or energetics. For a domain wall in Fig.~\ref{fig:domain-wall}, $\mathbf M \to \pm \mathcal M \hat{\mathbf z}$ as $y \to \pm \infty$ and $g = 2 \mathcal J$. If the domains were reversed, $\mathbf M \to \mp \mathcal M \hat{\mathbf z}$ as $y \to \pm \infty$, we would obtain $g = - 2 \mathcal J$. In terms of the $\mathbb Z_2$ topological charge $\zeta = \pm 1$ defined in Eq.~(\ref{eq:Z2-charge}), $g = 2 \zeta J$.

\subsection{Mass density and surface tension}

We consider a straight domain wall $y(x) = Y$ of length $\ell$ with the unit vector of magnetization $\mathbf m \equiv \mathbf M / \mathcal M$ interpolating between $\mathbf m(y) = (0,0, +1)$ and $(0,0,-1)$. In what follows, we parametrize the unit vector $\mathbf m = (\sin \theta \cos \phi, \sin \theta \sin \phi, \cos \theta)$ in terms of its polar and azimuthal angles $\theta$ and $\phi$. Potential energy consists of exchange
\begin{equation}
\begin{split}
U_{\text{exchange}} & 
	= A h \int d^2 \mathbf r \; \partial_i \textbf m \cdot \partial_i \mathbf m
    \\ 
& = A h \ell \int dy 
	\left[ 
    	\left(\partial_y \theta \right)^2 
        + \sin^2\theta \, 
        	\left(\partial_y \phi\right)^2 
    \right],
\end{split}
\end{equation}
easy-axis anisotropy
\begin{equation}
U_{\text{anisotropy}} = - K h \int d^2 \mathbf r \; m_z^2 = - K h \ell \int dy \; \cos^2 \theta,
\end{equation}
and the interfacial Dzyaloshinskii-Moriya interaction \cite{EPL.100.57002}
\begin{equation}
\begin{split}
U_{\text{DMI}} & = D h \int d^2 \mathbf r \;( m_z \nabla \cdot \textbf m -  \textbf m \cdot \nabla m_z)\\
& = D h \ell \int dy \; 
	\left(
    	\sin{\phi} \, \partial_y \theta 
        + \frac{1}{2} \sin{2\theta} \cos{\phi} \,
        	\partial_y \phi
    \right).
\end{split}
\end{equation}

Minimization of the energy with respect to the field $\phi(y)$ yields solutions with uniform $\phi(y) = \Phi_0 = \pm \pi/2$. For these, the energy as a functional of $\theta(y)$ is
\begin{equation}
U = 
	h \ell \int dy 
    	\left[ 
        	A \left(\frac{d\theta}{dy}\right)^2
            - K \cos^2{\theta}
            + D \sin \Phi_0\frac{d\theta}{dy}
        \right].
\label{eq:U-theta-y}
\end{equation}
The Dzyaloshinskii-Moriya term in Eq.~(\ref{eq:U-theta-y}) is topological: it does not depend on the exact profile $\theta(y)$ but only on the values of $\theta$ at $y = \pm\infty$. Therefore $\theta(y)$ is determined by minimization of the exchange and anisotropy energies, which yields

\begin{equation}
\cos\theta(y) = \zeta \tanh{\frac{y-Y}{\lambda}},
\label{eq:domain-wall}
\end{equation}
where $\lambda = \sqrt{A/K}$ is the width of the domain wall and 
\begin{equation}
\zeta \equiv 
	\frac{1}{2} \int_{-\infty}^\infty dy \, 
    	\frac{\partial \cos{\theta}}{\partial y}
     = \pm 1
\label{eq:Z2-charge}
\end{equation}
is the $\mathbb Z_2$ topological charge of the domain wall. The energy of the domain wall $U = h \ell (4 \sqrt{AK} - \pi \zeta D \sin{\Phi_0})$ is minimized when $\Phi_0 = \zeta \pi/2$, assuming $D>0$. That yields surface tension\cite{HeideDMI2008}
\begin{equation}
\sigma = h(4 \sqrt{AK} - \pi D). 
\label{eq:sigma-DMI}
\end{equation}

For small deviations from the N{\'e}el configuration, the energy varies quadratically in $\delta \Phi = \Phi - \Phi_0$,
\begin{equation}
U(\Phi) 
	= U(\Phi_0)
    + \frac{1}{2} h \ell \pi D \, (\Phi - \Phi_0)^2,
\end{equation}
giving the stiffness for in-plane magnetization $\kappa = \pi D h$. 

The Lagrangian for a domain wall of a general shape with slowly varying $Y(t,x)$ and $\Phi(t,x)$ is\cite{Makhfudz2012}
\begin{equation}
\mathcal L =
		g \dot{Y} \Phi - \frac{\sigma {Y'}^2}{2} - \frac{\kappa(\Phi - \Phi_0 - Y')^2}{2}.
\label{eq:L-Y-Phi}        
\end{equation}
The first term represents a gyroscopic coupling of strength $g = 2 \zeta \mathcal J$ between the position of the domain wall $Y$ and its azimuthal angle $\Phi$, \cite{Clarke2008} the second term comes from the expansion of the potential energy $U = \sigma \int dx \sqrt{1 + {Y'}^2}$ of a curved domain wall to the quadratic order in $Y'$, and the third term is the energy cost of a misalignment between in-plane magnetization and the direction of the domain wall. 

Minimization of the action $S = \int_\Omega dt \, dx \, \mathcal L$ with respect to the magnetization angle $\Phi$ yields 
\begin{equation}
\Phi = \Phi_0 + \frac{g}{\kappa} \dot{Y} + Y'.
\end{equation}
Eliminating $\Phi$ yields a Lagrangian for the field $Y(t,x)$, 
\begin{equation}
\mathcal L 
	= \frac{\rho \dot{Y}^2}{2} 
	- \frac{\sigma {Y'}^2}{2}
     + g \dot{Y} Y',
\end{equation}
with the D{\"o}ring mass density $\rho = g^2/\kappa$.\cite{Makhfudz2012} 

\subsection{Nonreciprocity}
\label{sec:nonreciprocity}

The dimensionless nonreciprocity parameter in the model of a N{\'e}el wall is 
\begin{equation}
\sinh{u} 
	= \frac{g}{\sqrt{\rho \sigma}}
    = \zeta \mathop{\mathrm{sgn}{\mathcal J}}
    	\sqrt{\frac{\pi D}{4 \sqrt{AK} - \pi D}},
\label{eq:u-D-A-K}
\end{equation}
As detailed in Sec.~\ref{sec:simulations}, $u$ can be of order 1 for realistic material parameters.

Coupling constants for a Bloch domain wall were derived in Ref.~\onlinecite{Makhfudz2012}. The nonreciprocity parameter in that model is $\sinh{u} = g/\sqrt{\rho \sigma} = 1/\sqrt{Q-1}$, where $Q = 2K/\mu_0 \mathcal M^2 > 1$ is a dimensionless measure of the easy-axis anisotropy $K$. In the ferromagnet FePt,\cite{Moutafis2009} $Q = 2.1$  yields strong nonreciprocity, $u = 0.85$.

\subsection{Range of applicability of the model}

The model of a nonreciprocal string represents a coarse-grained description of an actual domain wall and applies only on sufficiently large length scales and long times. Although the wave equation (\ref{eq:eom-y}) predicts sharp kinks on a domain wall (Fig.~\ref{fig:snapshots}), the kinks will be smooth below a certain length scale. We may reasonably guess that the characteristic length scale cannot be smaller than the width of the domain wall $\lambda$ and is possibly longer. Similarly, the linear relation $\omega = v_\pm k$ will break down at sufficiently high wavenumbers and frequencies and wave packets will exhibit dispersion. 

The following estimate was communicated to us recently by Kravchuk.\cite{Kravchuk} 

The minimal model of a domain wall (\ref{eq:L-Y-Phi}) neglects the increase of exchange energy associated with the gradient of the azimuthal angle $\Phi'$. Taking it into account produces the following Lagrangian:
\begin{equation}
\mathcal L = g \dot{Y} \Phi 
	- \frac{\sigma {Y'}^2}{2} 
    - \frac{\kappa(\Phi - \Phi_0 - Y')^2}{2}
    - \frac{(\sigma + \kappa) \lambda^2 {\Phi'}^2}{2}.
\end{equation}
After obtaining the equations of motion for the fields $Y(t,x)$ and $\Phi(t,x)$ in a standard way and performing a Fourier transform, we obtain the coupled equations for the amplitudes, 
\begin{equation}
\left(
	\begin{array}{cc}
    	(\sigma + \kappa)k^2 & - i g \omega + i \kappa k \\
        i g \omega - i \kappa k & \kappa + (\sigma + \kappa) \lambda^2 k^2
    \end{array}
\right)
\left(
	\begin{array}{c}
    	Y \\
        \Phi
    \end{array}
\right)
= 0,
\end{equation}
which yield the spectrum $\omega(k)$:
\begin{equation}
g\omega = \kappa k
	\pm k  (\sigma + \kappa) \sqrt {\frac{\kappa}{\sigma + \kappa} +  \lambda^2 k^2}.
\end{equation}
In the long-wavelength limit $k \to 0$, we recover the linear spectrum $\omega = v_\pm k$. (It helps to recall that $\rho = g^2/\kappa$.) This approximation is valid for wavenumbers
\begin{equation}
k \ll \lambda^{-1} \sqrt{\frac{\kappa}{\sigma + \kappa}}
	= \lambda^{-1} |\tanh{u}|.
\end{equation}
We thus find that, for moderate and strong nonreciprocity, $u \gtrsim 1$, our theory is indeed applicable on length scales exceeding the domain wall width $\lambda$. For weak nonreciprocity, $u \ll 1$, it only applies on much longer length scales exceeding $\lambda/u$. As Eq.~(\ref{eq:u-D-A-K}) shows, moderate nonreciprocity is achieved for substantially strong Dzyaloshinskii-Moriya coupling. Such values are achievable in thin ferromagnetic films. As we show in the next section, realistic material parameters used by \textcite{Boulle2013} yield $u$ of order 1. 

\section{Response to weak external perturbations}
\label{sec:response}

\subsection{General considerations}

With the motion of a free nonreciprocal string understood, we turn our attention to its dynamical response to weak external perturbations. They are mathematically represented by adding to the Lagrangian (\ref{eq:L-nonreciprocal}) terms that are linear in the displacement $y$ and its derivatives $\dot{y}$ and $y'$. We can start with the Lagrangian of a string coupled to a gauge field $a_\mu$ (\ref{eq:L-minimal-coupling}), which already contains terms $- \Lambda^{\mu\nu}a_\mu \partial_\nu y$, and add a term linear in $y$: 
\begin{equation}
\mathcal L 
	= \frac{1}{2} \Lambda^{\mu\nu} 
    	(\partial_\mu y - a_\mu)
        (\partial_\nu y - a_\nu) + p y.
\end{equation}
Here the source field $p(t,x)$ expresses pressure (force per unit length) exerted on the string by an external perturbation. The term quadratic in the gauge field does not contain $y$ and thus does not influence its dynamics.

Note that the pressure term $py$ breaks the translational symmetry in the $y$ direction and thus violates local conservation of the current (\ref{eq:j-gauge}), $\partial_\mu j^\mu = p$. Globally, transverse momentum 
\begin{equation}
P_y = \int_0^w dx \, j^0 
	= \int_0^w dx \, \Lambda^{0\nu} 
    	(\partial_\nu y - a_\nu)
\end{equation}
is no longer conserved: its rate of change is given by the net force from pressure:
\begin{equation}
\dot{P}_y = \int_0^w dx \, p.
\end{equation}

In the rest of this section, we derive the effects on the dynamics of a ferromagnetic domain wall of external perturbations such as an applied magnetic field and spin current. As anticipated above, their influences can be expressed in terms of pressure and emergent gauge potentials.

\subsection{Magnetic field perpendicular to the easy plane}

A magnetic field parallel to the hard axis, $\mathbf H = (0,0,H_z)$, breaks the energetic equivalence of the $\mathbf m = + \hat{\mathbf z}$ and $- \hat{\mathbf z}$ domains and thereby exerts pressure (force per unit length) 
\begin{equation}
p = -2 \zeta h \mu_0 \mathcal M H_z
\end{equation}
on a domain wall. Pressure couples directly to the field $y$, adding a term $\mathcal L_\text{ext} = py$ to the Lagrangian. The equation of motion (\ref{eq:eom-y}) changes to
\begin{equation}
\rho \ddot{y} - \sigma y'' + 2 g \dot{y}' = p.
\end{equation}
Because the Lagrangian becomes $y$-dependent, transverse momentum $P_y$ (\ref{eq:Py-def}) is no longer conserved and changes at a rate proportionally to the external force. 

When a uniform field is applied to a domain wall in equilibrium, $y = 0$, its momentum begins to increase at the rate $\dot{P_y} = p w$, producing the time-averaged acceleration $\langle \ddot{y} \rangle_t = p /\rho_\text{eff}$. The response of the non-periodic, momentum-carrying mode is
\begin{equation}
y_0(t,x) = \frac{p}{2 \rho_\text{eff}}
	\left[
    	t + \frac{g}{\sigma} 
        	\left(x-\frac{w}{2}\right)
    \right]^2.
\end{equation}

In the presence of dissipation, the string accelerates until it reaches a steady state, in which $\dot{P}_y = p w - b w \dot{y} = 0$. This determines the terminal velocity 
\begin{equation}
\lim_{t \to \infty}\dot{y}(x,t) = \frac{p}{b} 
	= -\zeta \frac{|\gamma| \mu_0 H_z \lambda}{\alpha}.
\end{equation}
Because	 the boundary condition (\ref{eq:bc-y}) is unchanged, the steady-state motion is a zero mode (\ref{eq:zero-mode}) with a constant slope $y' = gp/(b\sigma)$.

\subsection{Magnetic field parallel to the easy plane}

A magnetic field applied in the easy plane, $\mathbf H = (H_x, H_y, 0)$, couples to the in-plane components of magnetization $m_x$ and $m_y$, which exist only in the vicinity of the domain wall. The Zeeman coupling adds the following term to the Lagrangian of the string expressed in terms of fields $y$ and $\phi$ (\ref{eq:L-Y-Phi}):
\begin{equation}
\mathcal L_\text{ext}
	= \pi \lambda h \mu_0 \mathcal M 
		(H_x \cos{\phi} + H_y \sin{\phi}).
\end{equation}
To the first order in deviation $\delta \phi = \phi - \phi_0$ from an equilibrium state for a Bloch wall with $\phi_0 = 0$ or $\pi$ and for a N{\'e}el wall with $\phi_0 = \zeta \pi/2$, we obtain $\mathcal L_\text{ext} = \tau \, \delta \phi$, where $\tau$ is the torque per unit length,
\begin{equation}
\begin{aligned}
&\tau =  \pi \cos{\phi_0} \, \lambda h \mu_0 \mathcal M H_y 
\quad
&& \mbox{Bloch wall},
\\
&\tau = - \pi \sin{\phi_0} \, \lambda h \mu_0 \mathcal M H_x 
&& \mbox{N{\'e}el wall}.
\label{eq:torque}
\end{aligned}
\end{equation}

We ignore here the second order term $(\delta \phi)^2$, which will modify the torque in Eq.(\ref{eq:torque}) and give rise to corrections to the metric tensor $\Lambda^{\mu\nu}$.  

After integrating out the field $\phi$, we obtain the Lagrangian for the field $y$ alone:
\begin{equation}
\mathcal L =  
		\frac{\rho \dot{y}^2}{2} 
    	- \frac{\sigma {y'}^2}{2} 
    	+ g \dot{y} y'
    	+ \tau \left(y' + \rho\dot{y}/g\right).
\label{eq:L-transverse-field}
\end{equation}
Normally, Lagrangian terms linear in $\dot{y}$ and $y'$ do not affect the dynamics because their contribution to action $S = \int_\Omega dt \, dx \, \mathcal L$ reduces to boundary terms that drop out of the equations of motion. But not quite so here. If the external field is time-dependent then it will generate a force proportional to its time derivative: 
\begin{equation}
\rho \ddot{y} - \sigma y'' + 2 g \dot{y}' 
	= - \rho \dot{\tau}/g.
\label{eq:eom-torque}
\end{equation}
This force is particularly important when an external field is turned on or off, effectively giving the string a kick. 

The $y'$ term also contributes, albeit indirectly. It alters the boundary condition (\ref{eq:bc-y}):
\begin{equation}
\sigma y' - g \dot{y} = \tau
\mbox{ for }
x = 0, w.
\label{eq:bc-torque}
\end{equation}

The in-plane field preserves the translational symmetry in the $y$ direction. The transverse momentum acquires a contribution proportional to the in-plane field: 
\begin{equation}
P_y  = \int_0^w dx 
	\left(
    	\rho \dot{y} + g y' + \rho \tau/g
    \right),
\label{eq:Py-H}
\end{equation}
from which we can determine the zero mode velocity:
\begin{equation}
\langle \dot{y} \rangle_t = \frac{P_y}{M_\text{eff}} - \frac{\tau}{g},
\end{equation}
where the total amount of applied torque $\int_0^w dx \, \tau(t,x)$ contributes to the changing rate of the angular momentum $- g \int_0^w dx \, \dot{y}$.

Starting with a domain wall in equilibrium, $y(t,x) = 0$, we suddenly turn on an in-plane magnetic field at $t=0$. The transverse momentum is conserved and remains unchanged at $P_y = 0$. The domain wall keeps its horizontal orientation, $y'=0$, and picks up a constant velocity: 
\begin{equation}
y(t,x) = -\frac{\tau}{g} \, t \, \Theta(t),
\label{eq:y-turn-on-torque}
\end{equation}
where $\Theta(t)$ is the Heaviside step function. When dissipation is present, the domain wall eventually reaches zero velocity and slope $y' = \tau/\sigma$.

Alternatively, we can start with a  domain wall tilted in an external field and turn off the field at $t=0$. Again, only the zero mode is engaged:
\begin{equation}
y(t,x) = \frac{\tau}{\sigma} x 
	+ \frac{\tau}{g} \, t \, \Theta(t).
\end{equation}

It is worth noting that a sudden switching on or off of an easy-plane magnetic field generates a simple response of the domain wall that involves strictly the zero mode (\ref{eq:zero-mode}) but none of the oscillatory normal modes (\ref{eq:normal-modes}). Put differently, no kinks emerge from the edges when the field is switched on or off. The reason for this simple behavior becomes clear when we view the perturbation as a coupling to a background gauge field as described in Sec.~\ref{subsec:gauge}.

The torque term in the Lagrangian (\ref{eq:L-transverse-field}),
\begin{equation}
\mathcal L_\text{ext} 
	= \tau(y'+\rho \dot{y}/g) 
	= - \Lambda^{\mu\nu} a_\mu \partial_\nu y,
\end{equation}
corresponds to a gauge field with only a temporal component, 
\begin{equation}
a_0 = - \tau/g, 
\quad
a_1 = 0.
\end{equation}
If the torque $\tau(t)$ is time-dependent but spatially uniform then $a_\mu$ is a pure gauge, $\partial_0 a_1 - \partial_1 a_0 = 0$, and thus can be represented by a gradient, 
\begin{equation}
a_\mu = \partial_\mu \chi(t), 
\quad 
\chi(t) = -\frac{1}{g} \int dt \, \tau(t).
\end{equation}
With the aid of Eq.~(\ref{eq:response-to-pure-gauge}), we obtain the response
\begin{equation}
y(t,x) = -\frac{1}{g} \int dt \, \tau(t).
\end{equation}
This yields Eq.~(\ref{eq:y-turn-on-torque}) for a torque that is suddenly turned on at $t=0$.

\subsection{Spin transfer torque}

An electric current density along the transverse direction $j \hat{\mathbf y}$ injects an angular momentum of $-g\int_0^w dx \, v$ per unit time, where $v = \hbar P j /(2 q \mathcal J)$ is the electron drift velocity, $P$ is the spin polarization, and $q$ is the electron charge. The adiabatic torque density is $\tau = -g v$, and the non-adiabatic torque exerts pressure $p = |g| \beta v /\lambda$.\cite{Thiaville2005} In a steady state, 
\begin{equation}
y(t,x) = \frac{\beta }{\alpha} v t 
	+ \frac{\beta - \alpha}{\alpha}
    \frac{g v}{\sigma} x.
\end{equation} 

\begin{figure}[b]
\includegraphics[width = 1\linewidth]{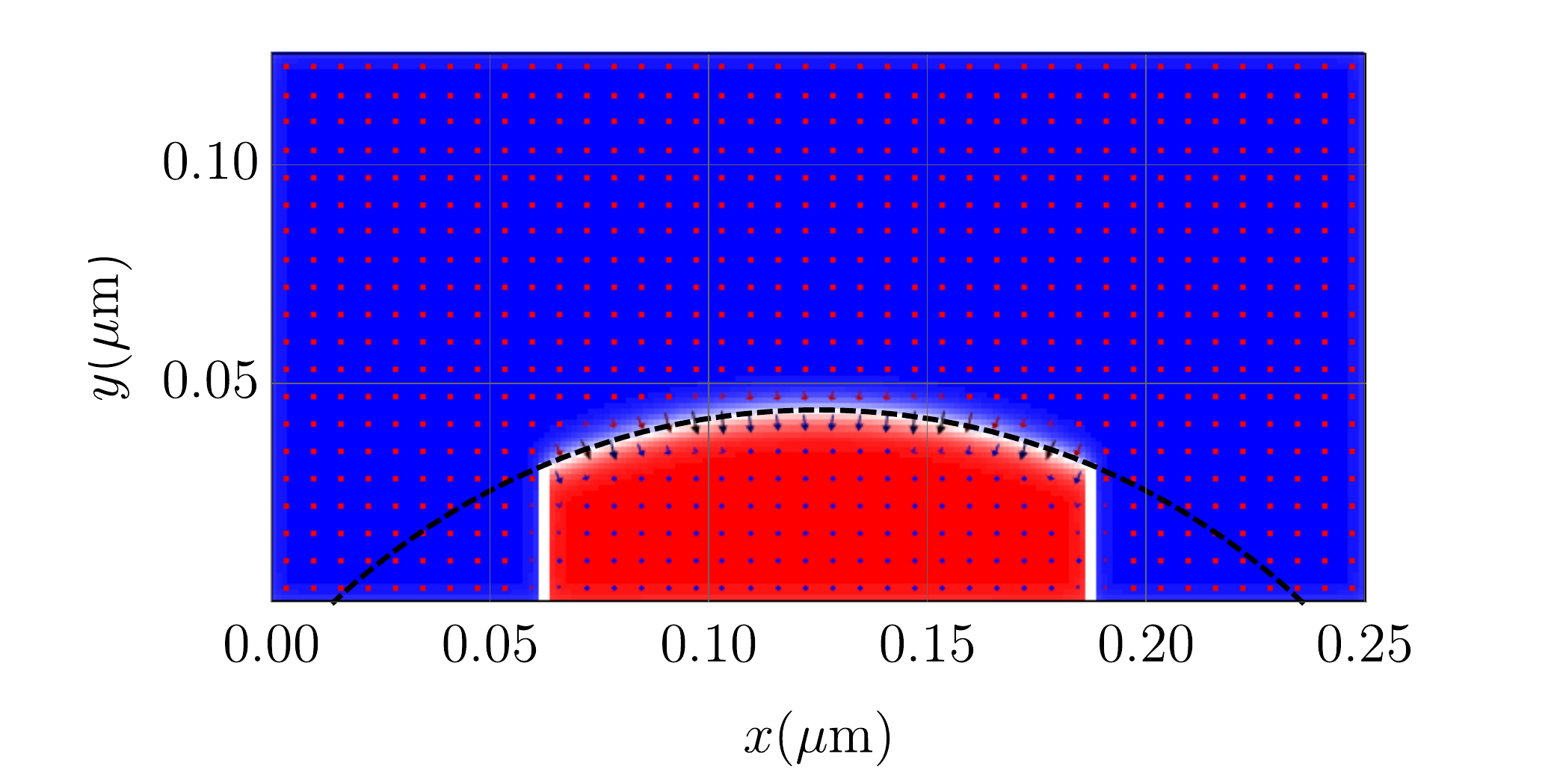}
\caption{Tension measurement. A domain wall pinned by lattice defects (white narrow vertical notches) and subject to an external magnetic field of strength $\mu_0 H = 25$ mT applied along the easy axis. The shape of the domain wall is a circle (black dashed line) of radius $R$}.
\label{fig:tension-measurement}
\end{figure}

\begin{figure}[b!]
\includegraphics[width = 1\linewidth]{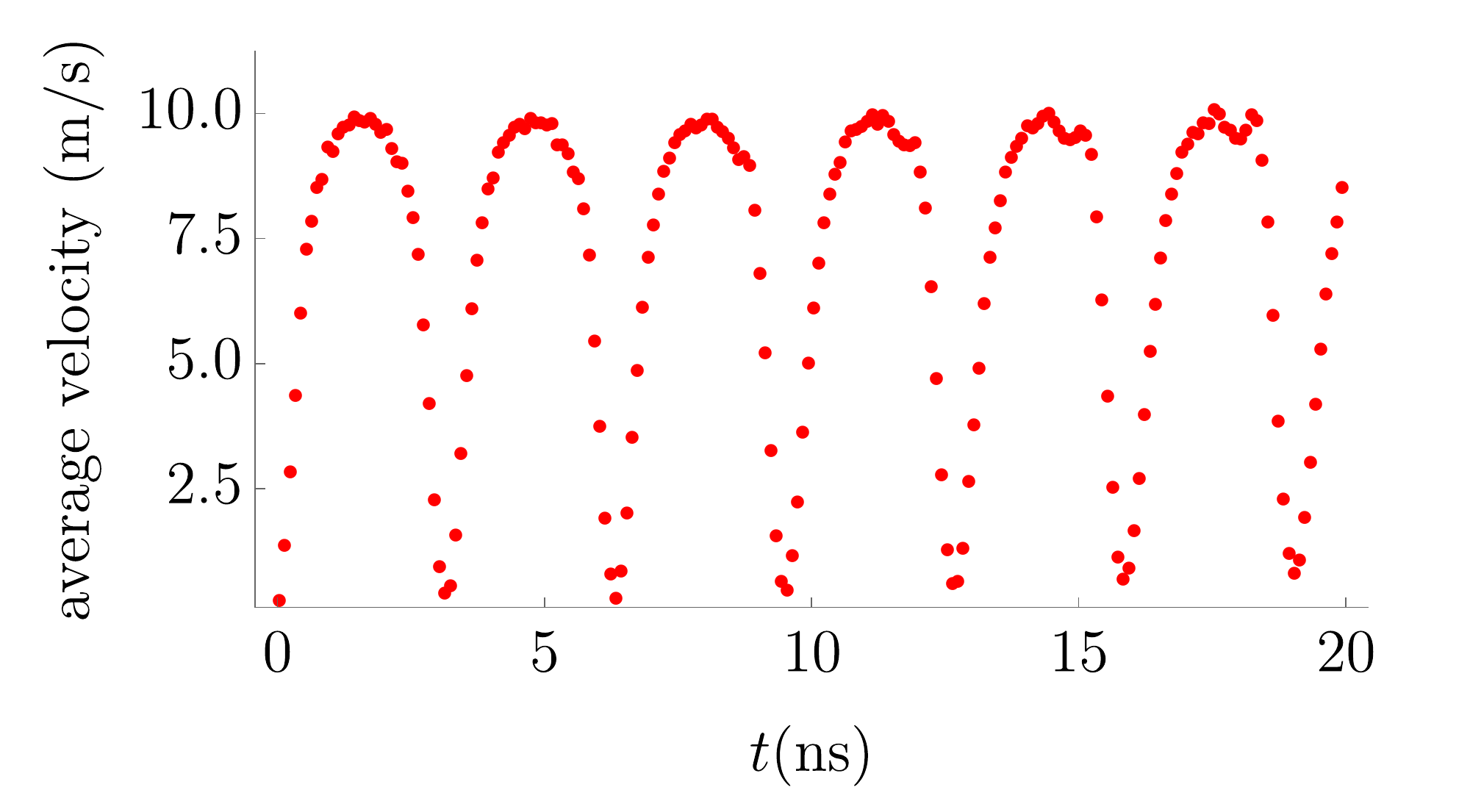}
\caption{Measurement of the fundamental period $T$ from the space-averaged velocity $\langle\dot{y}(t)\rangle_x$ averaged over the length of the domain wall. }
\label{fig:frequency-measurement}
\end{figure}

\section{Micromagnetic simulations}
\label{sec:simulations}

To check the accuracy of our model, we performed micromagnetic simulations in mumax$^3$ (Ref.~\onlinecite{mumax}) of a N{\'e}el domain wall stabilized by strong Dzyaloshinskii-Moriya interaction.\cite{EPL.100.57002} We used material parameters similar to those of \nolink{\citeauthor{Boulle2013}},\cite{Boulle2013} which yield the gyroscopic constant $g = 1.43 \times 10^{-14}$ J s/m$^2$, tension $\sigma = 1.02 \times 10^{-11}$ J/m, mass density $\rho = 2.58 \times 10^{-17}$ kg/m, characteristic speed (\ref{eq:c}) $c = 473$ m/s, and nonreciprocity $u = \mathop{\mathrm{arcsinh}{\frac{g}{\sqrt{\rho \sigma}}}} = 0.793$. 

The value for string tension $\sigma$ was measured in the simulation by pinning a straight domain wall at two points and applying a magnetic field along the easy axis to exert pressure $p_H = 2 \mu_0 \mathcal M H h$ on the domain wall. Under pressure, the wall deforms into a circular arc until the field pressure is balanced by the Laplace pressure $p_\sigma = \sigma /R$, where $R$ is the radius of the arc (Fig.~\ref{fig:tension-measurement}). We thus obtained $\sigma = 2 \mu_0 \mathcal M H R h  = 1.01 \times 10^{-11}$ J/m, in good agreement with the theoretical estimate. 

The characteristic speed $c = 472$ m/s, extracted from the oscillation period $T = 2w/c$ of the space-averaged transverse velocity 
\begin{equation}
\langle \dot{y}(t) \rangle_x \equiv \frac{1}{w} \int_0^w dx \, \dot{y}(t,x)
\end{equation}
of the domain wall (Fig.~\ref{fig:frequency-measurement}), was also in good agreement with the theoretical value (\ref{eq:c}). The corresponding mass density is $\rho = 2.56 \times 10^{-17}$ kg/m. 

Having thus determined and verified all independent parameters of the nonreciprocal string, we ran dynamical micromagnetic simulations to check the predictions of the string model against the simulated micromagnetic dynamics. A domain wall with an initial tilt 
\begin{equation}
a = \frac{y(0,w) - y(0,0)}{w} = 0.023
\end{equation}
is expected to have the time-averaged velocity (\ref{eq:average-velocity-kinks}) $\langle \dot{y} \rangle_t = c a \tanh u = 7.13$ m/s, in good agreement with the velocity of $7.11$ m/s observed in our simulations. 

Snapshots of the domain wall from the simulation are shown as dots in the right column of Fig.~\ref{fig:snapshots} against the theoretical shape (see video of the simulation in Supplemental Material \cite{suppmat}). The model describes the observed motion of the domain wall very well without any adjustable parameters. 

See Appendix \ref{app:simulation} for further technical details.

\section{Discussion}
\label{sec:discussion}

We have presented a simple theory of an extended domain wall in a thin-film ferromagnet wherein a domain wall is modeled as a string with nonreciprocal dynamics. The three parameters of the string---mass density $\rho$, surface tension $\sigma$, and the gyroscopic constant $g$---are closely related to material parameters or measurable in experiments. 

The nonreciprocity is directly related to the spontaneous breaking of the time-reversal symmetry in a ferromagnet. It is manifested in unequal speeds for transverse waves propagating on the string in opposite directions (\ref{eq:v}) and is quantified by a dimensionless parameter $\sinh{u} = g/\sqrt{\rho \sigma}$. It  gives rise to unusual dynamics: strange-looking normal modes in finite geometry, steady-state motion with a tilt, and more. Our estimates show that domain walls in thin ferromagnetic films with realistic material parameters can exhibit strong nonreciprocity $u$ of order 1 and a large disparity of the wave velocities, $v_+/v_- = - e^{2u}$.

The model has been tested against numerical simulations. It reproduces very well the dynamics of a N{\'e}el domain wall in a ferromagnet with Dzyaloshinskii-Moriya interaction.\cite{Boulle2013}  The model is also expected to describe the dynamics of Bloch domain walls.\cite{Makhfudz2012}

We have considered a nonreciprocal string under the simplifying assumptions of translational invariance in the transverse direction and of small deviations from a straight-line equilibrium shape. In realistic thin films, the presence of inhomogeneities requires an extension to arbitrary curves. This can be done by using the mathematical language of interface evolution developed previously for dissipative dynamics. \cite{PhysRevA.29.1335, PhysRevA.46.4894} It would also be interesting to expand the theory from two to three spatial dimensions, where a domain wall would become a two-dimensional surface.

Another restriction implicit in our analysis is the slowness of the dynamics. Fast motion will be accompanied by the generation and propagation of topological defects (Bloch lines \cite{Malozemoff1979}) within a domain wall. Inclusion of Bloch lines as stable point-like defects can be done along the lines of \nolink{\citeauthor{Nikiforov1986}}.\cite{Nikiforov1986} 

Last but not least, it would be interesting to look for other string-like objects with nonreciprocal dynamics outside magnetism.

\section*{Acknowledgments}

We thank Sayak Dasgupta, Se Kwon Kim, Volodymyr Kravchuk, Masaki Oshikawa, Jonathan Robbins, Valery Slastikov, and Oleg Tretiakov for helpful discussions. This work was supported by the US Department of Energy, Office of Basic Energy Sciences, Division of Materials Sciences and Engineering under Award DE-FG02-08ER46544.

\appendix

\section{General solution}
\label{app:Laplace}

We apply Laplace transform
\begin{equation}
\tilde{y} (s,x) = \int_0^\infty  y(t,x) e^{-st} dt.
\end{equation}
The equation of motion (\ref{eq:eom-y}) is transformed to
\begin{equation}
 - \sigma \tilde{y}''+ 2g s \tilde{y}' + \rho s^2 \tilde{y} = \mu(x).
 \label{eq:eom-y'}
\end{equation}
In the absence of external perturbations, $\mu(x) = \rho s y(0,x) + \rho \dot{y}(0,x) + 2g y'(0,x)$ is fully determined by the initial configurations. It is easy to find a particular solution $\tilde{y}_p(s,x)$ for the second-order differential equation (\ref{eq:eom-y'}) of a single variable $x$.  The full solution is thus $\tilde{y}(s,x) = \tilde{y}_p(s,x) + \tilde{y}_c(s,x)$, where the complementary solution $\tilde{y}_c(s,x) = \tilde{A}_+(s) e^{-s x /v_+} + \tilde{A}_-(s) e^{-sx/v_-}$ contains all the traveling modes.

Coefficients in $\tilde{y}_c(s,x)$ can be determined by the following boundary conditions:
\begin{equation}
\sigma \tilde{y}'_c - gs \tilde{y}_c = \tilde{\nu}_{0,1} (s)  \text{ for }  x= 0,w,
\end{equation}
where $\tilde{\nu}_{0} (s) = - g y(0,0) - \sigma \tilde{y}_p'(s,0) + g s \tilde{y}_p (s,0)$ and $\tilde{\nu}_{1} (s) = - g y(0,w) - \sigma \tilde{y}_p'(s,w) + g s \tilde{y}_p (s,w)$. 

In the example of a tilted-and-release string, $\tilde{y}_p(s,x) = ax/s$, $\tilde{\nu}_0 (s) = \tilde{\nu}_1 (s)= -\sigma a /s$. The inverse Laplace transform gives $y_p(t,x) = ax \,\Theta(t)$ and $\nu_0 (t) = \nu_1 (t) = - \sigma a \, \Theta(t)$, where $\Theta(t)$ is the Heaviside step function. 

Generally, the boundary conditions yield
\begin{equation}
\begin{split}
& \tilde{A}_+(s) =  \frac{\sigma}{c} \frac{e^{-sw/v_-}\tilde{\nu}_0 (s) - \tilde{\nu}_1(s)}{s(e^{-sw/v_+} - e^{-sw/v_-})},\\
& \tilde{A}_-(s) = \frac{\sigma}{c} \frac{e^{-sw/v_+}\tilde{\nu}_0 (s)- \tilde{\nu}_1(s)}{s(e^{-sw/v_+} - e^{-sw/v_-})},
\end{split}
\end{equation}
where we used $\sigma/ v_\pm + g = \pm \sigma/c$. The inverse Laplace transform can be then performed with the aid of the following series of step functions. We define
\begin{equation}
\begin{split}
& \alpha_0 (t) = \sum_n \Theta(t-nT),\\
& \alpha_1 (t) = \sum_n \Theta (t + w/v_- - nT ).
\end{split}
\end{equation}
The string evolution thus follows $y(t,x) = y_p(t,x) + A_+(t-x/v_+) \, \Theta(t-x/v_+) + A_-(t-x/v_-)$ for $t \geq 0$, where 
\begin{equation}
\begin{split}
A_+(t) = \frac{\sigma}{c} \int_0^t dt' & \big[ -\alpha_0(t') \nu_0(t-t') + \alpha_1 (t') \nu_1(t-t') \big], \\
A_-(t) = \frac{\sigma}{c}  \int_0^t dt' &  \big\{ [ \Theta(t')-\alpha_0(t') ] \nu_0(t-t') \\ 
& \qquad \qquad \qquad + \alpha_1 (t') \nu_1(t-t')  \big\},
\end{split}
\label{eq:time-convolution}
\end{equation}
and $\Theta(t-x/v_-) = 1$ was used.

External perturbations can then be readily built into $\mu(x)$ and $\nu_{0,1}(s)$ and enter into the time convolution (\ref{eq:time-convolution}).

\section{Simulation details}
\label{app:simulation}

Micromagnetic simulations were performed using the numerical package mumax$^3$. \cite{mumax} The sample was a strip of width $w = 1.50$ $\mu$m and thickness $h = 1.25$ nm with cell size $1.25$ nm. Material parameters similar to those of \nolink{\citeauthor{Boulle2013}}:\cite{Boulle2013} saturation magnetization $\mathcal{M} = 1.00 \times 10^6$ A/m, gyromagnetic ratio $\gamma = - 1.76 \times 10^{11} $ A s/kg, exchange stiffness $A = 1.00 \times 10^{-11}$ J/m, easy-axis anisotropy $K = 1.30 \times 10^{6}$ J/m$^3$, and Dzyaloshinskii-Moriya interaction $D = 2.00 \times 10^{-3}$ J/m$^2$. These yield the gyroscopic constant $g = - 2 \mathcal{M} h /\gamma = 1.43 \times 10^{-14}$ J s/m$^2$, domain-wall width $\lambda = 2.77$ nm, tension $\sigma = h(4 \sqrt{AK} - \pi D) = 1.02 \times 10^{-11}$ J/m, mass density $\rho = g^2/(\pi Dh) = 2.58 \times 10^{-17}$ kg/m, and nonreciprocity $u = \mathrm{arcsinh} (g/\sqrt{\rho \sigma}) = 0.793$. Long-range dipolar interactions were turned off. (Their local part can be taken into account by renormalization of the easy-axis anisotropy from $K$ to $K - \mu_0 \mathcal M^2/2$.) The sample was relaxed and then run with Gilbert damping $\alpha = 0$.

The model of a nonreciprocal string has three independent parameters: mass density $\rho$, surface tension $\sigma$, and the gyroscopic constant $g$. The gyroscopic constant is directly related to the material parameters: $g =  2 \zeta \mathcal J =  2\zeta \mathcal M h/\gamma$. To test our theory, we measured directly the surface tension $\sigma$ and the characteristic speed $c$ (\ref{eq:c}). 

The measurement of surface tension is illustrated in Fig.~\ref{fig:tension-measurement}. A domain wall was pinned at two ends by defects in the sample. A magnetic field of strength $\mu_0 H = 25.0$ mT was applied along the easy axis, creating pressure on the domain wall $p_H = 2 \mu_0 \mathcal M H h$. After relaxation, the domain wall attained the shape of a circular arc with the radius $R = 0.155 \; \mu$m. In equilibrium, the pressure from the magnetic field is balanced by the Laplace pressure of the curved domain wall $p_\sigma = \sigma/R$. This directly yields $\sigma = 2 \mu_0 \mathcal M H R h = 1.01 \times 10^{-11}$ J/m, in good agreement with the theoretical value.

To determine the D{\"o}ring mass density $\rho$, we measured the fundamental period of oscillations $T$ (\ref{eq:T}), which is related through the speed $c$ (\ref{eq:c}) to the string parameters: 
\begin{equation}
\frac{T}{2 w} = \frac{1}{c}
	= \sqrt{\frac{g^2}{\sigma^2}+\frac{\rho}{\sigma}}.
\end{equation}
In Fig.~\ref{fig:frequency-measurement}, we tracked the the evolution of a tilted domain wall over several periods and plotted the velocity averaged over the length of the domain wall as a function of time. The period $T = 3.18$ ns in a strip with the width $w = 0.75 \mu $m yielded $\rho = 2.56 \times 10^{-17}$ kg/m.

\begin{figure}[b]
\includegraphics[width = 0.6\linewidth]{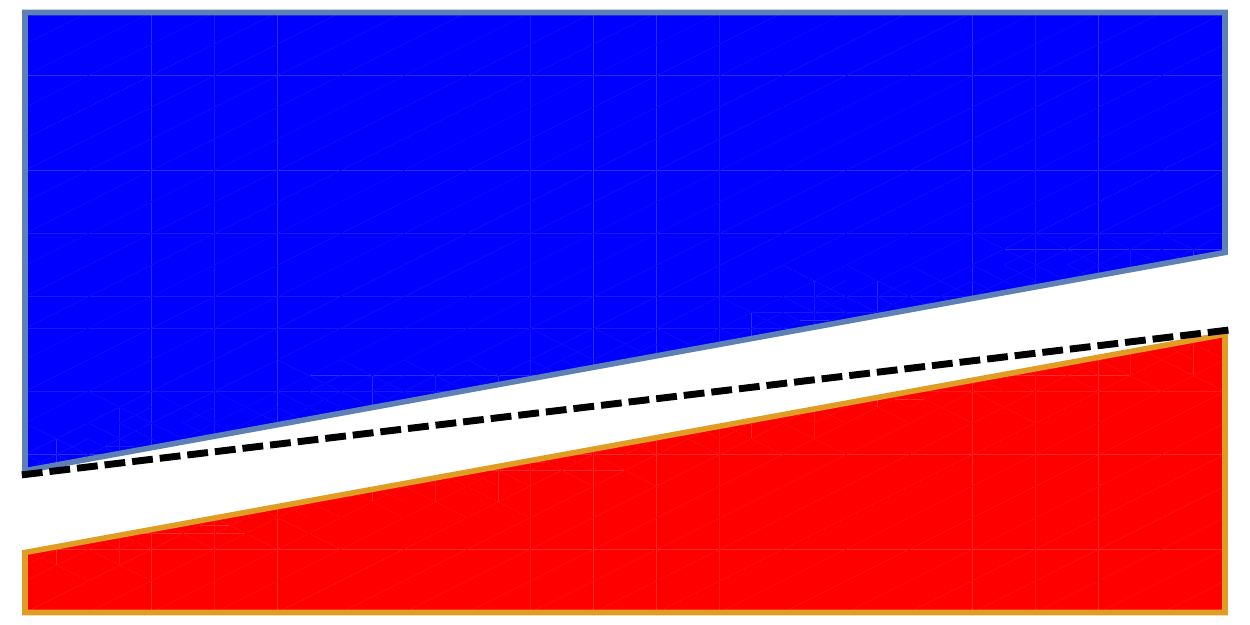}
\caption{Distribution of the out-of-plane magnetic field to create a tilted domain wall. In the blue and red regions, the applied field is parallel to the direction of local magnetization; in the white strip, whose width greatly exceeds the domain wall width $\lambda$, the applied field is zero. The domain wall relaxes to the dashed line under the combined effects of the applied field and tension.}
\label{fig:initial_tilt}
\end{figure}

\begin{figure}[b!]
\includegraphics[width = 0.75\linewidth]{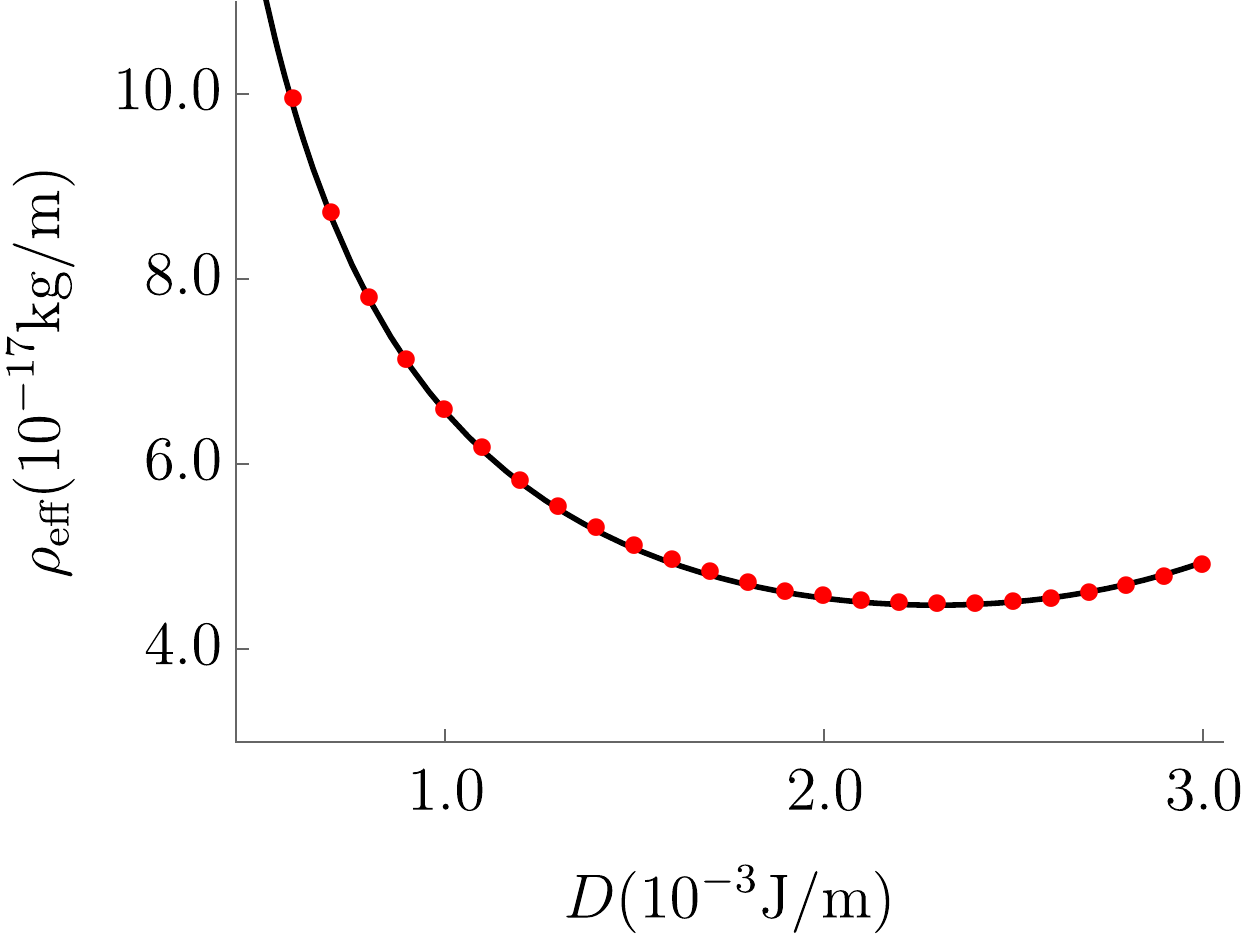}
\caption{The dependence of effective mass density on the DMI in the model (black line) and the simulation (red dots).}
\label{fig:effective_mass}
\end{figure}

To prepare a tilted domain wall with zero initial velocity, we used a nonuniform out-of-plane magnetic field schematically shown in Fig.~\ref{fig:initial_tilt} to restrict the initial relaxation of the domain wall. The field was then set to zero to release the domain wall when the evolution began.

As a further check of the phenomenological model, we measured the effective mass density $\rho_\mathrm{eff}$ as a function of the Dzyaloshinskii-Moriya coupling $D$. Our theory relates the effective mass density to the initial slope $a$ of the domain wall and its velocity $\left\langle \dot{y} \right\rangle_t$ averaged over a period $T$:
\begin{equation}
\rho_\mathrm{eff} = \frac{ga}{\left\langle \dot{y} \right\rangle_t}.
\end{equation}
 The effective mass density extracted in this way is plotted as a function of $D$ in Fig.~\ref{fig:effective_mass}. It shows excellent agreement with the expected functional dependence 
\begin{equation}
\rho_\mathrm{eff} 
	= \rho + \frac{g^2}{\sigma}
	= 4 \mathcal J^2 
    	\left[
        	\frac{1}{\pi D h} 
            + \frac{1}{(4 \sqrt{AK}-\pi D)h}
        \right].
\end{equation}

We also provide videos to show the dynamics. Videos 1-3 present the normal modes of a string under fixed and free boundary conditions for different values of nonreciprocity $u$;  Video 4 shows the time evolution of tilted-and-released strings also for different $u$ values; and Video 5 compares the prediction of the nonreciprocal string model to the simulation result of a tilted-and-released domain wall with simulation parameters given here.

\bibliographystyle{apsrev4-1}
\bibliography{string}

\end{document}